\documentclass[twocolumn,epsfig,showpacs,floatfix,superscriptaddress,preprintnumbers,amsmath,amssymb]{revtex4}
\usepackage{amsmath}
\usepackage{amssymb}
\usepackage{graphicx,psfrag,subfigure}
\usepackage{dcolumn}
\usepackage{bm}

\def\prb{Phys. Rev. B}
\def\prl{Phys. Rev. Lett.}

\def\be{\begin{equation}}
\def\ee{\end{equation}}
\def\ba{\begin{eqnarray}}
\def\ea{\end{eqnarray}}

\def\C60{A$_x$C$_{60}$}

\begin{document}

\title{Quasi-1D dynamics and nematic phases in the 2D Emery model}
\author{Steven A. Kivelson}
\affiliation{Department of Physics, University of California at Los Angeles,  Los 
Angeles, CA 90095}
\author{Eduardo Fradkin}
\affiliation{Department of Physics, University of Illinois, 1110 W. Green Street, 
Urbana, IL 61801-3080}
\author{Theodore H. Geballe}
\affiliation{Department of Applied Physics, Stanford University, Stanford, CA. 
94305-4045}
\date{\today}

\begin{abstract}
We consider the Emery model of a 
 Cu-O plane of the high temperature
superconductors.  We show that in a strong-coupling limit, with strong Coulomb 
repulsions between
electrons on nearest-neighbor O sites, the electron-dynamics is strictly one 
dimensional, and
consequently a number of asymptotically exact results can be obtained concerning the
electronic  structure.  
In particular, we show that a nematic phase, which spontaneously breaks the point-
group symmetry 
of the square lattice, is stable at low enough temperatures and strong enough 
coupling.    
\end{abstract}

\maketitle

Immediately following the discovery of high temperature superconductivity in the 
cuprates, 
it was realized \cite{pwa87,vje87} that the novel physics of these materials is 
dominated by 
the strong, short-range repulsion between electrons.  However, there has been 
considerable 
debate over what is the simplest ``paradigmatic" model that captures the 
essential physics 
of the problem.
Despite the fact, pointed out early on by Emery \cite{vje87} and by Varma and
coworkers \cite{varma87},
that  the minimal 
model which captures the essential local chemistry of the doped copper-oxide 
planes is the 
three-band copper-oxide or Emery model (defined below),  
it has generally been the 
accepted practice among theoreticians to, instead, consider the single-band 
Hubbard or t-J 
model - certainly reasonable models for studying the 
interplay 
between the localized quantum antiferromagnetism of the undoped system and the 
charge 
delocalization produced by doping.    
Moreover, since none of these strongly interacting models 
can be solved in two dimensions (2D), any theoretical
results that can be established with an acceptable degree of rigor can
shed light on the  observed physics of 
the actual materials.  

In this paper we show that there exists a limit (which is not wildly  
unphysical) of 
the Emery 
model about which a number of exact statements are possible.  Specifically, 
despite the fact 
that the model itself possesses the symmetries of the square lattice, 
the electron 
dynamics 
is quasi-one-dimensional in this limit.  It is also possible to establish the 
existence of 
various electronic liquid crystalline phases \cite{kfe99}, including especially 
an Ising-
nematic phase \cite{pokrovsky} which spontaneously breaks the four-fold rotational 
symmetry
of  the 
underlying lattice.  

\section{The model}
\label{sec:model}

We consider a model defined on the copper-oxide lattice, shown in 
Fig. \ref{lattice};  the corresponding Hamiltonian operator
is written explicitly in Appendix A.  The  copper sites 
define a simple square lattice at lattice positions $\vec R$ with lattice 
constant, $a$, while 
the oxygen sites sit at the center of the nearest-neighbor bonds on this
lattice,  and so define a 
second square lattice, rotated by 45$^\circ$ relative to the Cu lattice, with 
lattice positions 
$\vec R + (a/2)\hat e_x$ and $\vec R + (a/2)\hat e_y$, which we will call site 
$(\vec R,x)$ 
and $(\vec R,y)$, respectively. 
The vacuum is defined as the 
state in which all the O p orbitals and Cu d-orbitals 
are full.  The relevant Fock space is constructed by adding holes (removing 
electrons) from 
the Cu 
3d$_{x^2-y^2}$ and O 2p$_{\sigma}$ orbitals ({\it i.e.} the 2p$_x$ orbital 
associated with the oxygens at $\vec R,x$ and the 2p$_y$ for the oxygens at 
$\vec 
R,y$.)  
The corresponding hole creation operators are $d_{\vec R,\sigma}^{\dagger}$ and 
$p_{\vec R,a,\sigma}^{\dagger}$.

\begin{figure}[bht]
\psfrag{ud}{$U_{d}$}
\psfrag{up}{$U_{p}$}
\psfrag{vpd}{$V_{pd}$}
\psfrag{vpp}{$V_{pp}$}
\psfrag{tpp}{$t_{pp}$}
\psfrag{tpd}{$t_{pd}$}
\psfrag{eps}{$\epsilon$}
\begin{center}
\includegraphics[width=0.4\textwidth]{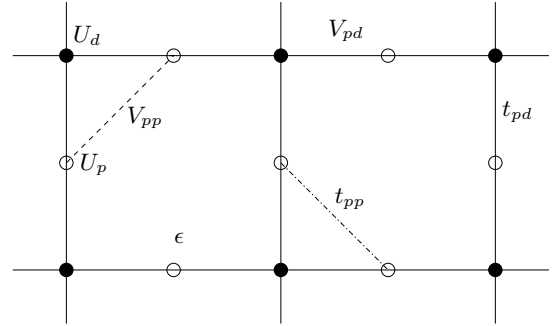} 
\end{center}
\caption
{Schematic representation of the $Cu-O$ lattice.  
The full circles represent $Cu$ sites and the
open circles are $O$ sites.  
The various terms in the  Emery-model Hamiltonian are represented, as 
discussed in the text.}
\label{lattice}
\end{figure}

The various interactions in the model (also shown in Fig. \ref{lattice}) 
are defined as follows:  
The repulsion
between two holes on the same site is $U_d$ and $U_p$, respectively, 
for a copper and oxygen 
site, while
the repulsion between two holes on an adjacent copper and oxygen  
or a nearest-neighbor pair 
of oxygens are
$V_{pd}$ and $V_{pp}$.  All further neighbor interactions are neglected.  
The hopping matrix 
elements
which transfer a hole between a nearest-neighbor O - Cu pair is 
$t_{pd}\equiv t$, while that 
between
nearest-neighbor O's is $t_{pp}$, and the difference  between the energy 
of an electron on 
an O
and Cu site is $\epsilon > 0$.  The signs of the various hopping matrix 
elements are 
determined by the
symmetry of  the 
relevant d and p orbitals. However,  a simple gauge transformation with wave-
vector $\vec 
\pi \equiv (\pi/a)\left( 1,1 \right)$,  changes the signs so that all 
the relevant 
hopping matrix elements are positive. 

The insulating parent state of the undoped cuprates
has one hole per unit cell which, because $\epsilon >0$, live
preferentially  on the Cu sites.
Additional doped holes, whose concentration per unit cell we denote $x$,
go  preferentially on O
sites because $U_{d}\gg \epsilon$.

\begin{figure}[bht]
\psfrag{a}{a)}
\psfrag{b}{b)}
\psfrag{c}{c)}
\psfrag{d}{d)}
\psfrag{e}{e)}
\begin{center}
\includegraphics[width=0.4\textwidth]{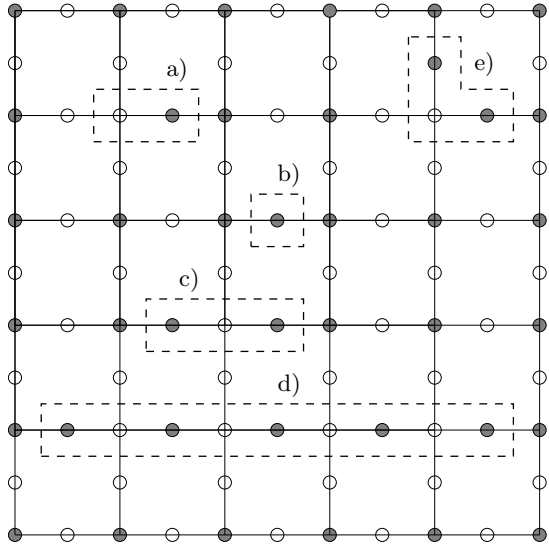} 
\end{center}
\caption
{Schematic representation of various states discussed in the text. The dark
circles are $Cu$ and $O$ sites occupied by a hole; the open circles are 
sites not occupied by holes.}
\label{states}
\end{figure}

\section{The Strong Coupling Limit}  
\label{sec:strong-coupling}

We start by defining the ``strong-coupling limit'' of this model, in which  the 
interaction strengths are large compared to
the one-electron energies.  Here we discuss the salient features of this regime. 
In Appendix \ref{app:strong} we present details and prove that the statements we make 
here are asymptotically exact in this limit.

By the strong coupling limit we formally mean that we consider the model in the 
limit $U/t \to \infty$, where
all the interactions, $U_p, U_d, V_{pd},
V_{pp} \sim U$ and the hopping matrix elements, $t_{pd}\equiv t$ and $ t_{pp} \sim 
t$.   Even in this limit, the physics
depends on the finite ratios of the various interaction strengths.  In particular, 
we will always assume that the following
inequalities are satisfied:  $U_d > \epsilon >0$, and
$U_d > U_p > V_{pd} >V_{pp} >0$, 
consistent with chemical intuition.  (Somewhat more restrictive inequalities must be
assumed in order to prove all the stated results, as is discussed explicitly
in Appendix A.)  Finally, since the hopping  matrix
elements depend exponentially on separation, we set $t_{pp}/t\to 0$.  This final 
assumption may not be well satisfied in 
the actual materials, where cluster calculations \cite{cluster} suggest that 
$t_{pp}/t \sim 1/3$. We will study this model
as a function of 
$x$ and for arbitrary ratio of $\epsilon/t$.

For the undoped system, $x=0$, the ground-state has zero energy and is $2^N$ fold 
degenerate, with one hole on each copper.  Of course, this degeneracy is
resolved  for finite 
interaction strengths when antiferromagnetic super-exchange interactions, with  
$J\approx 8t^4/U_p V_{pd}^2$, are included. However, in the 
strong-coupling limit, this (and most of
the other spin-physics we will encounter) involves energy scales that vanish as 
$t/U$ and $t/V \to 0$;  we will therefore ignore this physics at first, and 
then return to it when we consider
``$t-J$-like" physics that arises from low order corrections to the 
strong-coupling limit.  

Neglecting
the spin degeneracy, the first excited state is  an  exciton, shown in Fig.
\ref{states}a, with a large energy
$E_{ex}= V_{pd} +
\epsilon+ {\cal  O}(t)$, 
and so can be ignored at low energies and temperatures.

Now, consider one additional doped hole.  Since $U\gg V$, adding one hole means 
increasing the number of occupied sites by 1, and so necessarily costs a 
minimum energy of 
$\mu\equiv 2V_{pd}+\epsilon$.   Some possible representative states are
shown in Figs.  \ref{states}b-d:  
\begin{itemize}
\item
\ref{states}b shows the bare hole-state, 
which to zeroth
order in $t$   has energy $\mu$,  
\item
\ref{states}c  shows a hole-exciton bound 
state, with
zeroth order energy $\mu+\epsilon$.
\item
 \ref{states}d shows a hole  broken 
into two
charge $e/2$ solitons separated by $L=4$ sites, with zeroth order energy 
$\mu+L\epsilon$.
\item
\ref{states}e shows a bent hole-exciton bound-state which is
the lowest energy state which involves a disturbance 
outside of 
this row and   which 
has zeroth 
order 
energy $\mu+V_{pp}+\epsilon$, and so can be neglected in the strong coupling 
limit \cite{chris}.  
\end{itemize}

The most salient point to notice is that  if the doped hole 
is added to 
an oxygen on a given row, all states with energy near $\mu$ involve disturbances 
which are  confined to the same row.  Any state which involves a non-colinear 
disturbance,
such as the bent-hole exciton in \ref{states}e, costs infinite energy in the 
strong
coupling limit.

In this limit, therefore,  the number of  holes on each row and each column  
of the 
lattice are 
separately conserved quantities, and the charge dynamics is purely 
one-dimensional.  More precisely, in Appendix \ref{app:strong} we show that, to 
leading order in the strong coupling expansion and for $t_{pp}=0$, each row $p$ 
has a conserved quantity $X_p$, and each column $q$ a conserved quantity 
$Y_q$, which qualitatively correspond the number of hole
quasiparticles  on that row or column.
Indeed, doped holes
on distinct parallel rows do not interact with each other. However holes on rows 
and columns do interact 
with each other (where they meet). 
We 
will show below  that these interactions play a crucial role. Of course,  when we 
back off 
from the strong
coupling limit, or if we include a small but non-zero $t_{pp}$, small effective 
interactions 
which
violate these conditions will be generated.

\section{The 1D dynamics}
\label{sec:1D}

Consider a system in which we add a fixed number of doped holes to one and only 
one row of the lattice.  Because
the number of electrons in each row is conserved, none of these holes can leak out 
onto other rows or columns, which thus
remain undoped.  Indeed, the electron dynamics along  this row is  exactly 
equivalent to
those of the 1D Cu-O model, which was analyzed previously in 
Ref. \cite{emery1D}:
\ba
H_{row}=&&-t\sum_{j,\sigma}[c^{\dagger}_{j,\sigma}c_{j+1,\sigma}+ h.c.] + 
\sum_j \epsilon_j\hat n_j \nonumber \\
&&+\sum_j[U_j\hat n_{j,\uparrow}\hat n_{j,\downarrow} + 
V_{pd}\hat n_j\hat n_{j+1}]
\label{Hrow}
\ea
where even number sites are Cu sites and odd numbered sites are O, 
$\epsilon_{2j}=0$,
$\epsilon_{2j+1}=\epsilon$, $U_{2j}=U_d$, 
$U_{2j+1}=U_p$, $\hat n_{j,\sigma}=c^{\dagger}_{j,\sigma}c_{j,\sigma}$, and 
$\hat n_j=\sum_{\sigma}
\hat n_{j,\sigma}$.  In the strong coupling $U_j\to\infty$ limit, the charge 
degrees of freedom can be
treated as  spinless fermions~\cite{emery1D},
with effective Hamiltonian
\begin{equation}
H_c=-t\sum_{j}[c^{\dagger}_{j}c_{j+1}+ h.c.] + 
\sum_j[\epsilon_j\hat n_j+V_{pd}\hat n_j\hat n_{j+1}],
\label{Hc}
\end{equation}
while again the dynamics of the spin-degrees of freedom are obtained only when
corrections to the strong coupling limit of order $x t^2/U_j$ are included.
The density of spinless fermions per site is simply 
$(2N)^{-1}\sum_j\hat n_j=1+x$.

Manifestly, for $x=0$, the system is insulating, with one hole on each 
copper site and a 
charge gap
$\Delta_c=2 V_{pd}+{\cal O}(t)$.  For small, positive $x$, the 
{\it doped holes} are 
dilute and can be treated within the
context of an effective mass approximation, as free spinless fermions with 
creation energy 
$\Delta_c$ and
effective mass,
$m^*$. In particular, the ground state energy per site (with $W \equiv 
\hbar^2\pi^2/6m^*$)is 
\begin{equation}
E=E_0+\Delta_c \; x + W x^3 +
{\cal O}(x^5).   
\label{fermipressure}
\end{equation}
Both $\Delta_c$ and $W$ are continuous functions of $\epsilon/t$:
$\Delta_c=2V_{pd}+\epsilon F_{\Delta}(\epsilon/t)$ and $W = (8\pi^2/3) t a^2 
F_W(\epsilon/t)$.
For
$\epsilon\gg t$,
$F_{\Delta}=1+{\cal O}(t/\epsilon)^2$ and $W = 8 (t/\epsilon)[1+{\cal 
O}(t/\epsilon)^2]$. 
In the opposite limit, 
$\epsilon/t \to 0^+$, the 
fermions
fractionalize to form twice as many charge $e/2$ solitonic Fermions.  
However, the ground state 
energy has
the same $x$ dependence, but with $F_{\Delta}=-4(t/\epsilon)[ 1+{\cal 
O}(\epsilon/t)]$ and 
$W = 1+{\cal O}(\epsilon/t)$.  An important qualitative point to recognize here is 
that
the Fermi pressure is a decreasing function of $\epsilon/t$ which
vanishes  as $\epsilon/t\to \infty$.  Various
correlation functions can be accurately estimated, as well, from the well 
known theory of 
the 1D Luttinger liquid.  

For $x=1$, the system is again insulating, with one hole on each site.  
Expanding about 
this limit, for $1-x$
small, yields a result similar to those obtained 
for small $x$.
Other interesting states occur in the vicinity of various commensurate values 
of $x$.  
For instance, 
$x=1/2$ corresponds to commensurability 2 in the spinless fermion problem, 
where an incompressible
CDW state is the ground state of this strongly interacting problem. 

Indeed, given the large number of exact, or well controlled approximate 
results that can 
be obtained
for the 1DEG, and the ease with which quantum Monte-Carlo simulations can be 
employed to 
flesh out the
analytic results quantitatively, we consider the problem of a single, 
Cu-O row to be a solved problem.  This also means that a large number of 
``fully nematic'' states can also be completely characterized.  A fully
nematic state is defined to be one in which
which doped holes are placed only on rows (or only on columns).  Since holes
on neighboring rows do not interact at all (unless we were to add longer range 
interactions 
to the model), the dynamics of the holes on each row are determined by the same 1D 
Hamiltonian
we have just analyzed.  This does not constitute a complete solution of the 
problem, since  states
in which some doped holes lie on rows, and others on columns are still 
complicated, and require additional
analysis to characterize.  However, we will show below that, under many 
circumstances, the ground state is fully nematic.  

\section{The nematic phase} 
\label{sec:nematic}

We now move from the analysis of the hole 
dynamics 
along a single row or column, to
study the phases of the full two-dimensional model.
 
\subsection{The nematic insulator, $x=1$}
\label{sec:nematic-insulator}

At $x=1$ there are two holes per unit cell, 
and in the ground state
 each Cu and O site along each row is occupied by exactly one hole,
while the O sites along columns are empty. 
The energy per unit cell of
this state is $2V_{pd}+\epsilon$. There is a second, degenerate,
ground state obtained  by exchanging rows and columns.
Each ground state spontaneously breaks the the $90^\circ$ 
rotational invariance of the square
lattice but is translationally invariant since all unit
cells are equivalent. Furthermore these  strong
coupling ground states at $x=1$ have a charge gap. Hence this phase is a
{\sl nematic insulator}.   

\subsection{The ``classical limit," $\epsilon/t \to \infty $}
\label{sec:classical}

In this limit   
(still with $U_d>\epsilon$)
the  {\it charge} degrees of freedom define a classical lattice gas 
which can be mapped 
precisely to  an 
antiferromagnetic Ising model on the oxygen lattice with exchange coupling 
$V_{pp}/4$. Here spin-up indicates 
an occupied
state and spin down an unoccupied state.  
Under this mapping, the magnetization of the Ising 
model is $m=1-x$ and the N{\'e}el state is the insulating nematic state.  

\begin{figure}[bht]
\psfrag{I}{\large Isotropic}
\psfrag{N}{\large Nematic}
\psfrag{Q}{\large Quantum Nematic}
\psfrag{2}{\large 2 Phase}
\psfrag{t}{$T$}
\psfrag{tc}{$T_c$}
\psfrag{0}{0}
\psfrag{1}{1}
\psfrag{x}{$x$}
\begin{center}
\includegraphics[width=0.45\textwidth]{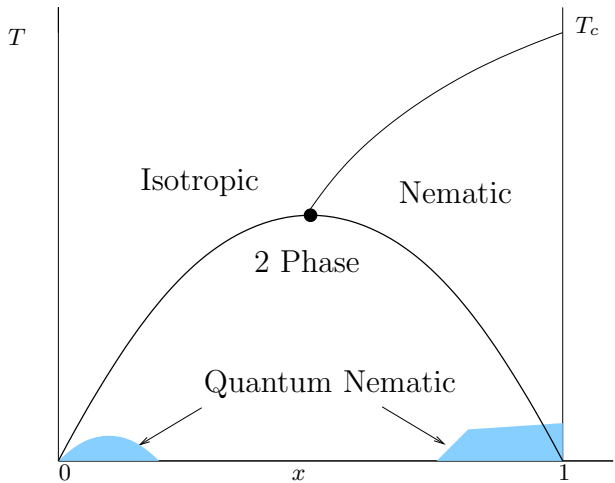} 
\end{center}
\caption
{Schematic phase diagram as a function of temperature and doped-hole 
concentration in the strong coupling limit 
(with magnetic ordering suppressed).  
Only the classical phase boundaries 
 are shown;
thick (thin) lines: discontinuous (continuous) transitions;
circle: tricritical
point; shaded regions: nematic phases stabilized at $t/\epsilon>0$.} 
\label{phase}
\end{figure}

The phase diagram of this model, shown in Fig. \ref{phase}, is well 
known~\cite{griffiths}. 
For $x\lesssim 1$, there is a continuous finite 
temperature
transition from a high temperature disordered phase to the low temperature 
N{\`e}el phase.    
However, at a critical
$x=x_c$, there is a tricritical point, such that for $x<x_c$ the
transition is discontinuous.  At low temperatures, for any $0 < x < 1$,
there is two-phase coexistence  between a
ferromagnetic ($x=0$ Mott-insulating) and a N{\`e}el 
($x=1$ insulating nematic) phase. 

At elevated temperatures, this classical phase diagram is relatively 
insensitive to the addition of perturbations to the Hamiltonian. 
Specifically, even for $t/\epsilon > 0$, the phase diagram is hardly 
altered by quantum effects so long as
$T \gg t / \sqrt{ t^2 + \epsilon^2 }$. However, at low temperatures, 
additional longer range interactions, either added explicitly to the 
model or induced by quantum fluctuations, can affect the nature of the 
stable phases substantially. For instance, even at the classical level, 
including the effect of weak Coulomb repulsion between holes on 
second-neighbor O orbitals will stabilize an electronic crystalline 
phase in a narrow range of $x$ near $1/2$ and at low enough temperatures. 
 Here the doped holes form a period 2 (Wigner crystal) density wave along 
 each row, 
while the O's along vertical bonds remain undoped. 
More generally, at low $T$ and intermediate $x$, the phase diagram is 
complex and dependent on details. However, for $x$ near 0 or 1, we 
will see that quantum effects generically stabilize homogeneous quantum 
nematic phases, shown as shaded areas in Fig. \ref{phase}.

\subsection{Quantum effects, $t / \epsilon > 0$, for $x \ll 1$ and $1-x \ll 1$}
\label{sec:quantum}

At low temperatures, $T \ll t / \sqrt{ t^2 + \epsilon^2 }$, quantum 
effects are important, even for small $ t / \epsilon$. We start by 
addressing the nature of the low temperature phase for 
$x \ll 1$.

Since the doped-holes are fermions, one would generally expect an 
associated Fermi pressure which tends to favor a uniform state. 
As shown in Eq.(\ref{fermipressure}), the 1D dynamics of the 
doped holes imply that the Fermi pressure $\sim W x^3$, rather than the 
stronger $W x^2$ dependence of a 2D Fermi gas. 
However, absent any attractive effective 
interactions between doped holes 
the Fermi pressure is always sufficient to stabilize a uniform 
phase at small $x$. 
(Effective attractions can arise from magnetic
fluctuations but are negligible at strong coupling.)

More specifically, an upper-bound to the ground state energy per site, 
$E(x)$, at all $x$ is given by the energy of the fully nematic state, 
$E_{nem}(x)$, {\it i.e.} the ground state energy of the 1D Emery model 
computed from Eq.(\ref{Hrow}) or Eq.(\ref{Hc}). It can easily be 
imagined that there is an isotropic state (or a less nematic nematic 
state) that has lower energy. However, while the properties of such 
an isotropic 
state cannot be computed exactly, a simple estimate shows that for 
$x\ll 1$, the ground
state is always nematic.  

To see this,
we compare the energy of the nematic phase with that of an isotropic 
version of this phase.  
If we ignore the interactions
between doped holes on rows and columns, the isotropic phase would have 
lower energy,  
since there are twice as
many 1D systems each with 1/2 the density of doped holes, resulting in a 
factor of 
4 reduction in the Fermi pressure. 
Now, for $x$ small enough, the contributions 
to the ground state energy from the $V_{pp}$  coupling between holes on crossing
Cu-O  rows and columns is a is regular function of $x$, free of infrared
divergences.
In particular, when two doped holes approach each other at the intersection of a 
row and column, the repel each other
strongly, and the probability of such an interaction is proportional to $x^2$.  
Thus,
the energy per site of the nematic state is 
$E_{\rm nem}=E_0 +\Delta_c \; x + Wx^3 +
\ldots$, while that of the isotropic state is
$E_{\rm iso}=E_0 +\Delta_c \; x + (1/4)Wx^3 +V^{\rm eff} x^2 +\ldots$ where
$V^{\rm eff}$  is an effective repulsion
between holes on intersecting rows and columns.  
Manifestly, at small enough $x$, the nematic state has lower energy.

The Fermi pressure similarly stabilizes the uniform phase for 
$x\lesssim 1$. Here the nematic character of the resulting uniform phase 
is considerably more obvious:  
the nematic phase consists of an array of Luttinger liquids
(one per row) with an effective Luttinger charge parameter
$K_c$ and an effective charge velocity $v_c$. 

\section{Corrections to the Strong Coupling limit}  
\label{sec:corrections}

Low order corrections to the strong coupling limit 
 resolve the spin-degeneracies we have neglected, 
and produce other
important changes in the physics, which we will
discuss elsewhere; here 
we  comment on a
few salient features.

\subsection{Magnetic Interactions}
\label{magnetic}

In the undoped Mott insulator at $x=0$, the most obvious induced
interaction \cite{vje87} is the  super-exchange interaction between
nearest-neighbor Cu spins, which to leading (4$^{th}$) order in $t$ is
$J=8t^4/[(V_{pd}+\epsilon)^2(U_p+2\epsilon)]\{1+(U_p+2\epsilon)/(2U_d)\}$. 
However, unlike the one-band model, under some circumstances,  for the strong 
coupling limit of the Emery model other
higher order interactions can be comparable in magnitude to this
interaction.  Thus, there is a 4-spin ring exchange interaction
on a plaquette generated at 8$^{th}$ order, which  does not
vanish for $U\to\infty$.  (For $U_d=U_d=\infty$ and
$V_{pd}=V_{pp}=V\gg\epsilon$, we get 
$J_4= (23/2)t^8/V^7$ while the leading order contribution to $J$ is
$J\sim t^4t_{pp}/V^4$.)  Thus, whereas the Mott insulating ground state at
$x=0$ of the single band model is inevitably magnetically
ordered, the Emery model has, in addition, quantum disordered (likely
dimerized) phases.

\subsection{2D Charge Dynamics}
\label{2D-charge}

For any of the conducting phases discussed above, the corrections to
strong coupling not only resolve the spin degeneracies, but also lead to
important changes in the charge dynamics at asymptotically low
energies.  In particular, under most circumstances, we expect that the
peculiar non-Fermi liquid behavior resulting from the strictly 1D
dynamics of the strong coupling limit will be destroyed at vanishing
temperatures by induced interactions (either proportional to $t_{pp}$ or
$t^2/V_{pp}$) which permit holes to hop from a row to a column.  These
couplings will either lead to a crossover to Fermi liquid behavior at low
temperatures, or to a broken symmetry
ground state.  However, so long as these interactions are weak, they
cannot restore the point group symmetry, so the nematic character of the
resulting states should be robust \cite{vadim,metzner}.  Moreover, as is 
characteristic of
quasi 1D systems, the non-Fermi liquid character will still be manifest
at non-zero temperatures, energies, or wave-vectors. 

\subsection{Coulomb interactions}
\label{sec:coulomb}

The fact that the model considered
has regions of two-phase coexistence, and others where various
susceptibilities (including the compressibility) are large, means that
the low temperature physics can be strongly modified by the effect of
even weak additional interactions.
Of these, the most obvious is the long-range Coulomb interaction, which
always frustrates phase separation and instead results in various locally
inhomogeneous phases such as stripe and bubble phases.

\subsection{Particle-Hole Asymmetry} 
\label{sec:particle-hole}

At least at strong coupling, 
the Mott insulating state at $x=0$ is strongly
particle-hole asymmetric: added electrons ($x < 0$) remove
holes from the Cu sites; these ``doped electrons"
have no local tendency to be dynamically confined to rows or columns.

\section{Conclusions}
\label{conclusions}

It has long been accepted that the Emery model provides a reasonable
description of the relevant electronic degrees of freedom in the
cuprates.  In this paper, we have obtained theoretically well controlled
results in a strong coupling regime of this model.  Although cluster
calculations suggest that this limit may not be entirely appropriate in
the real materials, the insights obtained here may  nevertheless capture
important features of the physics.

Perhaps the most salient feature of the results obtained here is the
existence of a strongly nematic phase in a significant portion of the
phase diagram.  This contrasts with the behavior of the same model in the
weak coupling limit, where it behaves in similar fashion to the single
band Hubbard model in which such a phase, if it occurs at all, is confined
to special fillings associated with the proximity to van Hove
singularities~\cite{metzner,kee}.  Experimental evidence of the existence
of a nematic phase~\cite{kfe99} in the cuprates was recently reviewed in
Ref. \cite{rmp}.  The other feature of the phase diagram is the existence
of a large region of two-phase coexistence - phase separation.  This may
be the simplest example of a generic tendency of highly correlated systems
to form inhomogeneous states~\cite{ek93}.

\section{Acknowledgements}
 
We thank B. Moyzhes and J. Tranquada for discussions.
This work was supported in part by
the National Science Foundation 
grants No. DMR 01-10329 (SAK), DMR 01-32990 
(EF), and by the Department of Energy contract DE-AC03-76SF00515 (THG).

\appendix
\section{Details of the derivations}
\label{app:strong}

In this Appendix, we formalize the statements concerning the strong coupling 
limit.  
The Cu sites are labeled by the Bravais lattice vectors, $\vec  R$, and the {\it 
hole} density on the Cu is $n({\vec R})$.  The
hole density on the two O sites in the same unit cell are labeled $n_{\pm x}(\vec R)$
and $n_{\pm y}(\vec R)$,
respectively.  (Notice that $n_{-x}(\vec R)=n_x(\vec R-a \hat x)$, and $n_{-y}(\vec R)=n_y(\vec R-a \hat y)$.) 
In the strong coupling limit, the Hamiltonian for the model described in Fig. \ref{lattice} can be written as the sum of two terms, $H=H_0+H_1$, with
\begin{widetext}
\begin{eqnarray}
H_0 &=&  \frac{U_d}{2} \sum_{\vec R} n({\vec R})\left( n({\vec R})-1\right) +  
\frac{U_p}{2} \sum_{{\vec R}} 
\left[n_x(\vec R) \left( n_x(\vec R)-1\right)+n_y(\vec R)\left(n_y(\vec R)-
1\right) \right]
\nonumber \\
&& +V_{pd}\sum_{\vec R}\left(  n({\vec R})-1\right) \left( n_x(\vec R)+n_{-x}(\vec 
R) + n_y(\vec R)+n_{-y}(\vec R)-2\right)
\nonumber\\
&& +V_{pp}\sum_{\vec R}  \left[n_x({\vec R}) n_y({\vec R})+n_y(\vec R)n_{-x}(\vec 
R)+n_{-x}(\vec R) n_{-y}(\vec R)+n_{-y}(\vec R)n_x(\vec R)\right]
\label{eq:H0}
\end{eqnarray}
and
\begin{eqnarray}
H_1&=&-t \sum_{\vec R,\sigma}
\left [
d^{\dagger}_{\vec R,\sigma} p_{\vec R,x,\sigma} +  
d^{\dagger}_{\vec R,\sigma} p_{\vec R-a\hat x,x,\sigma}+
d^{\dagger}_{\vec R,\sigma} p_{\vec R,y,\sigma}+
d^{\dagger}_{\vec R,\sigma} p_{\vec R-a\hat y,y,\sigma}+
{\rm h. c.}
\right] 
+\epsilon\sum_{{\vec R}} \left[n_x(\vec R)+n_y(\vec R)\right]
\label{eq:h1}
\end{eqnarray}
\end{widetext}
where $\sigma=\pm$ is the spin label.
As discussed in the text, we set direct O-O hopping amplitude $t_{pp}$ to zero.

To begin with we consider just the effect of the unperturbed Hamiltonian $H_0$. To make the proofs simpler, we will consider the
case in which  the following inequalities are satisfied:
 \begin{equation}
  U_d>2V_{pd}, \quad U_p>2V_{pd},\quad  V_{pp}>\frac{V_{pd}}{2}.
  \label{eq:ineq} 
  \end{equation}
In this case,  $H_0$ is positive semidefinite, as we will now demonstrate.  It is 
apparent from the structure of the Cu-O
lattice, that for every Cu site
$\vec R$ it is possible to define four triangles, located respectively NE, NW, SW 
and SE of the Cu site, and each having the Cu
site and two adjacent O sites for its vertices (see Fig. \ref{lattice}). We will 
label a triangle $\{\vec R,s,s^{\prime}\}$
according to its Cu vertex
$\vec R$ and by a pair of labels $s=\pm$ and $s^\prime=\pm$, where 
$\{s,s^\prime\}=\{+,+\}$ corresponds to the
triangle NE of $\vec R$,
$\{s,s^\prime\}=\{+,-\}$ to the triangle NW,  etc.
We denote by $n_{ss^\prime}(\vec R)$ the total hole occupancy of triangle  
$\{\vec R,s,s^\prime\}$:
\begin{equation}
n_{ss^\prime}(\vec R)=n(\vec R)+n_{sx}({\vec R})+n_{s^\prime y}(\vec R).
\label{eq:n-triangle}
\end{equation}
 By using this notation we can equivalently write $H_0$ in terms of the hole 
occupancy of each triangle and of the occupancy of the Cu and O sites:
 \begin{widetext}
 \begin{eqnarray}
 H_0&=&  \left(\frac{U_d}{2}-V_{pd}\right) \sum_{\vec R} n(\vec R)\left(n(\vec R)-
1\right)
 +\left(\frac{U_p}{2}-V_{pd}\right) \sum_{\vec R} \left\{n_x(\vec R)\left(n_x(\vec 
R)-1\right)+
 n_y(\vec R)\left(n_y(\vec R)-1\right)\right\}
 \nonumber \\
 &&+\frac{V_{pd}}{2}  \sum_{\vec R,s,s^\prime}\left(n_{ss^\prime}(\vec R)-
1\right)\left(n_{ss^\prime}(\vec R)-2\right)
 \nonumber  \\
&&+\left(V_{pp}-\frac{V_{pd}}{2} \right) \sum_{\vec R} 
\left\{n_x({\vec R}) n_y({\vec R})+n_y(\vec R)n_{-x}(\vec R)+n_{-x}(\vec R) n_{-
y}(\vec R)+n_{-y}(\vec R)n_x(\vec R)\right\}
 \nonumber \\
 &&
 \label{eq:H0-new}
 \end{eqnarray}
 \end{widetext}
Since each operator appearing in this expression is individually positive 
semidefinite, so is $H_0$, so long as all
 the  inequalities of Eq. \eqref{eq:ineq} are satisfied.

(Strictly, in the strong coupling limit, violating the inequalities of Eq. \ref{eq:ineq} can lead
to major restructuring of the ground state.  For instance, for 
$V_{pp}<V_{pd}/2$, the system phase  will separate for any
$x>0$, albeit if $t$ is not infinitesimal, quantum effects may lead to inhomogeneous
ordered  states. Indeed, close to the ``fully frustrated'' point
$V_{pp}=V_{pd}/2$, interesting forms of  quantum order-from-disorder effects
can arise. 
We will not consider these  interesting issues further in this paper. )

 Thus, provided the inequalities of Eq. \eqref{eq:ineq} are satisfied, the Hilbert 
space of (generally degenerate) zero energy states consists of the set of 
configurations in which, a) the Cu and the O sites are either empty or singly 
occupied (by holes), b) each triangle is either singly or doubly occupied (also by 
holes), and c) nearest-neighboring O sites are not simultaneously occupied by 
holes. All other states are separated from these zero energy states by a finite energy
gap. This is the low energy  Hilbert space of
states that we will consider here. 
 
 Hence, so long as $1\ge x \ge 0$, there exist zero energy ground states and an 
extensive ground-state
degeneracy.  Here, the density of ``doped holes,'' $x$, is defined in terms of the 
total hole density per site,
\begin{equation}
  1+x \equiv \frac{1}{N} \sum_{\vec R} \left\{  n(\vec R)+ n_x(\vec R)+n_y(\vec R) 
\ \right\}
  \label{eq:x}
\end{equation}
which is, of course, a conserved quantity.  ($N$ is the total number of Cu sites 
on this lattice.)

 To analyze the strong coupling limit to lowest order in perturbation theory, we 
confine our attention to the zero energy subspace of
the full Hilbert space.  Thus, the effective Hamiltonian, $H_{\rm eff}$, is 
obtained by
simply taking matrix elements of the perturbing Hamiltonian, $H_1$, defined in Eq. \eqref{eq:h1}, between states
in this subspace.
Higher order terms
in the perturbative expansion can be obtained, as is done in deriving the
$t-J$ model from the large
$U$ limit of the Hubbard model, by including the effects of virtual transitions to 
the finite energy states of the unperturbed
Hamiltonian. For the present purposes, we discuss only the first order problem, 
{\it i.e.\/}
\begin{equation}
H_{\rm eff}= P_0 H_1 P_0
\end{equation}
where $P_0$ is the projection operator onto the subspace of zero energy 
eigenstates
of $H_0$.  (See Fig. \ref{states}).

For $x=0$, there is a three-fold orbital degeneracy of the unperturbed ground 
state, in addition to the $2^N$-fold spin degeneracy
(which is only lifted, as discussed in the text, when high order superexchange 
processes are included).  Any $\epsilon >0$
eliminates this orbital degeneracy, and uniquely choses the ground-state with one 
hole on each Cu site;  there are no non-zero
matrix elements of the hopping term between states in the ground-state manifold.

For $x=1$, there is a two-fold orbital degeneracy of the ground state, as already 
discussed in the text.  As the two states
involved are related by a symmetry operation of the Hamiltonian (rotation by 
$\pi/4$) this degeneracy is not lifted in any order of
perturbation theory - this is the nematic insulating phase discussed in the text.  

For $\epsilon\gg t$, we perform a second perturbation expansion, which we refer to 
in the text as the ``classical'' limit:  To
first order in $\epsilon$, the ground state degeneracy is reduced to the subset of
states which have precisely
$xN$ holes  on O sites, {\it i.e.} the fewest number
possible. Since the term proportional to $t$ in $H_1$ changes the number of holes 
on O sites, there are no matrix elements between
these states to fist order in $t$.  However, to order $t^2/\epsilon$, it is 
possible for holes to move without violating this
constraint.

\begin{figure}[t]
\psfrag{1}{1}
\psfrag{2}{2}
\psfrag{3}{3}
\psfrag{4}{4}
\psfrag{to}{\Large $\!\!\! \Longrightarrow$}
\psfrag{a}{\small a)}
\psfrag{b}{\small b)}
\psfrag{c}{\small Cu}
\psfrag{O}{O}
\psfrag{p}{$p$}
\begin{center}
\subfigure{
\includegraphics[width=0.4\textwidth]{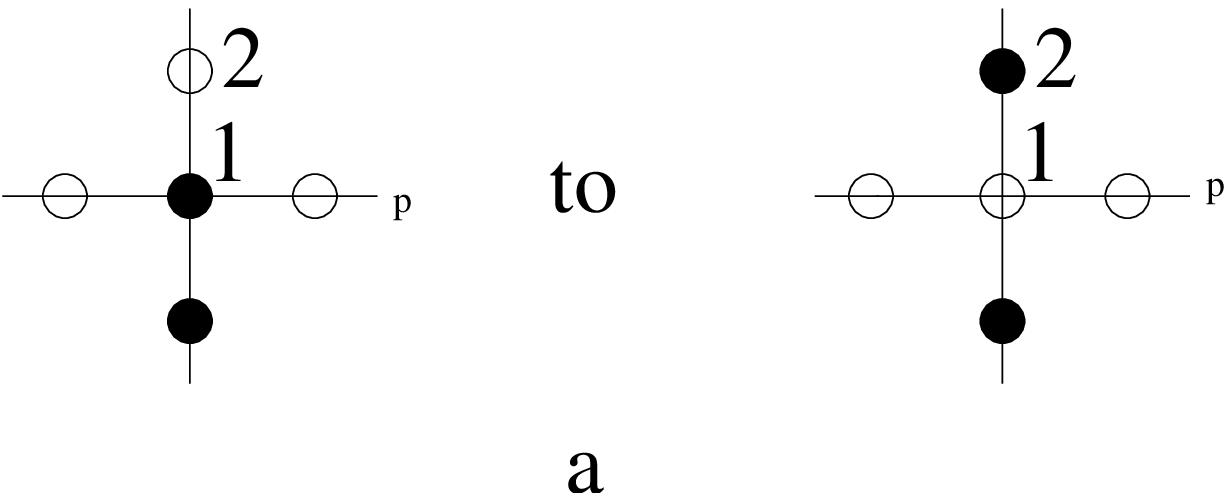}
}
\vskip 0.5cm
\subfigure{
\includegraphics[width=0.4\textwidth]{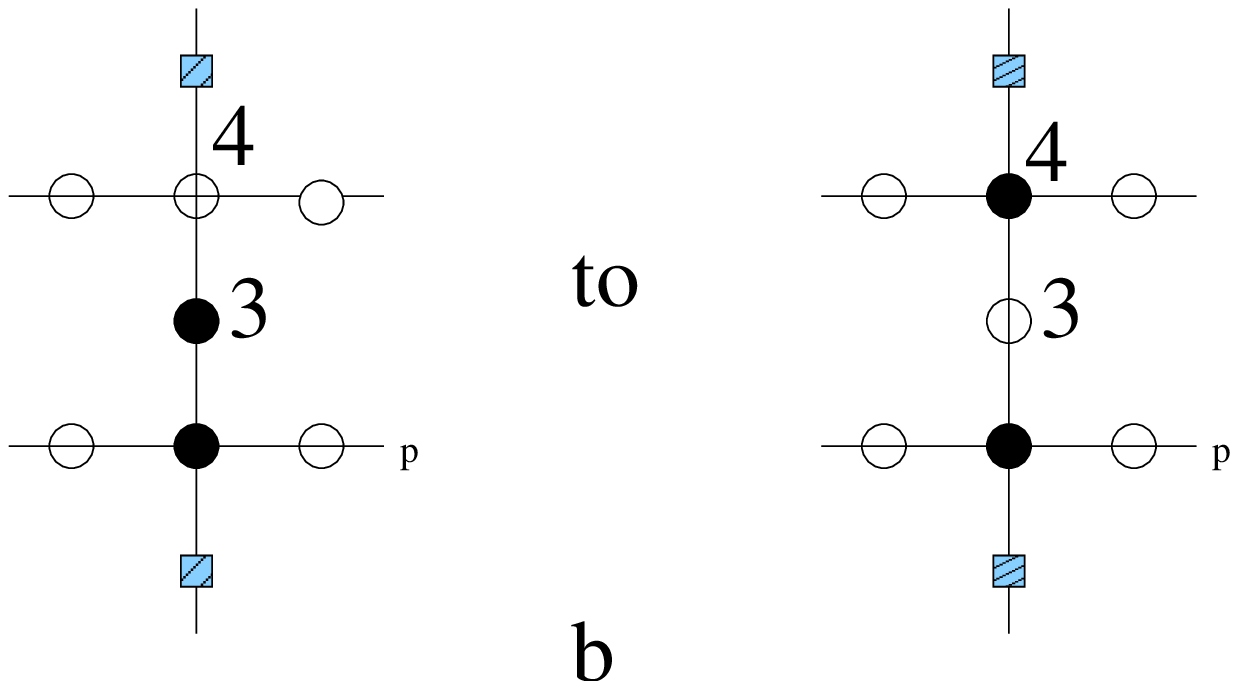}
}
\end{center}
\caption
{Conservation of $X_p$.  Shown, is a schematic of a segment of the $p^{th}$ row of 
Cu-O sites and its immediate neighborhood.  Cu sites are 
located at the intersections of the lines; O sites are half-way between two Cu 
sites. Filled (empty) circles are occupied (empty) sites. Empty squares are sites 
that can either be occupied or empty.
In a) we consider the change in the state produced when a hole hops between a Cu 
site labeled 1 in the row and the O site labeled 2.
In the initial state, the 
Cu site must be occupied by a hole and the  
O site must be empty.  In order that both the initial and final state survive 
projection with $P_0$, the O sites  
to the left and right must be empty and
the O site  
just below must be occupied.  Thus, this process necessarily decreases $n(pa\hat 
x)$ by 1, increases $N_x(pa\hat x)$ by 1, and
hence leaves $X_p$ unchanged.  
In b) we illustrate the same considerations for applied to the state in which a 
hole hops from the 
O site labeled 3 to the
Cu site labeled 4,
and the notation is the same. The state of the O sites with squares is not 
uniquely determined.
In this case, this process leaves unchanged $n(pa\hat x)=1$ and $N_x(pa\hat x)=0$,
and so does not change $X_p$.  Clearly,
the same considerations imply that the inverse processes, and the processes 
involving sites immediately below the $p^{th}$ row conserve $X_p$.  All
other processes in $H_1$ trivially commute with
$X_p$. }
\label{conserve}
\end{figure}

For intermediate values of $x$ and $t/\epsilon$, the effective Hamiltonian is generally
fairly  complicated and we have not obtained a general
solution.  However, we will now prove that under the dynamics of $H_{eff}$, there are 
conserved quantities, $X_p$ and $Y_p$,
corresponding to the number of ``quasiparticles'' in each row or column:
\begin{equation}
X_p\equiv \sum_{q}\left\{  n(q,p) + n_x(q,p)
+N_x(q,p)
\right\}
\label{row}
\end{equation}
where $(q,p)$ is the Cu site $qa\hat x + {pa}  \hat y$, and $N_x(\vec R)$ is an 
operator defined by
\begin{equation}
N_x(\vec R)=
\begin{cases}
1 & \textrm{if} \quad n_y(\vec R)=n_{-y}(\vec R)=1,\; n(\vec R)=0 \\
0 & \textrm{otherwise}
\end{cases}
\label{eq:Nx}
\end{equation}
Explicitly, since doubly occupied sites are anyway suppressed by $U_d$ and $U_p$, 
\begin{equation}
N_x(\vec R) = [1-n(\vec R)]n_y(\vec R)n_{-y}(\vec R).
\end{equation}
The first two terms in the sum in Eq. \eqref{row} count the number of holes
along the $p^{th}$ row.   
The third term, as we shall see, properly accounts for the finite transverse width 
of the actual quasi-particle excitations
by counting 
the number of vacant Cu sites along the row which have an occupied O site {\it
both} above and below.  The column operators, $Y_p$, and the associated operator 
$N_y(\vec R)$, are defined analogously.  

To show that each of these quantities is conserved, we first compute the 
commutator $[H_1,X_p]$ and then (since $[P_0,X_p]=0$) sandwich the resulting 
expression between projection operators. $X_p$ trivially commutes with all terms 
in $H_1$ except those that hop a hole between a Cu site in row $p$ and the O site 
immediately above (Fig. \ref{conserve}a) or below it, or between
a neighboring O sites and the next Cu site immediately beyond it (Fig. 
\ref{conserve}b.)  In general, the application of these terms changes
the value of $X_p$. However, as can be seen in the figure, and is explained in the 
caption, if we enforce the condition that the initial and final
state after the application of  $H_1$ is still a zero energy eigenstate of $H_0$, 
only processes which conserve $X_p$ survive.  This completes the
proof.

\end{document}